\documentstyle[aps,prl,twocolumn]{revtex}

\begin{document}

\twocolumn[\hsize\textwidth\columnwidth\hsize\csname %
@twocolumnfalse\endcsname

\title{Identification of Nuclear Relaxation Processes in a Gapped
Quantum Magnet: \\
$^{1}$H NMR in the $S=\frac{1}{2}$ Heisenberg Ladder
Cu$_2$(C$_5$H$_{12}$N$_2$)$_2$Cl$_4$}

\author{G. Chaboussant$^1$, M.-H. Julien$^1$, Y. Fagot-Revurat$^1$,
L.P. L\'evy$^{1,2}$, C. Berthier$^{1,3}$, M. Horvati\'c$^{1}$
and O. Piovesana$^{4}$}
\address{$^1$Grenoble High Magnetic Field Laboratory, CNRS and MPI-FKF,
BP 166, F-38042 Grenoble Cedex 9, France}
\address{$^2$Institut Universitaire de France and
Universit\'e J. Fourier, BP 41, F-38400 St. Martin d'H\`eres, France}
\address{$^3$Laboratoire de Spectrom\'etrie Physique, Universit\'e
J. Fourier, BP 87, F-38402 St. Martin d'H\`eres, France}
\address{$^4$Dipartimento di Chimica, Universit\`a di Perugia,
I-06100 Perugia, Italy}

\date{\today}
\maketitle

\widetext

\begin{abstract}

The $^1$H hyperfine shift $K$ and NMR relaxation rate T$_1^{-1}$ have
been measured as a function of temperature in the $S=1/2$
Heisenberg antiferromagnetic ladder Cu$_2$(C$_5$H$_{12}$N$_2$)$_2$Cl$_4$.
The presence of a spin gap $\Delta \simeq J_\perp-J_\|$ in this strongly
coupled ladder ($J_\| < J_\perp$) is supported by the $K$ {\em and} $T_1^{-1}$ results.
By comparing $T_1^{-1}$ at two different $^1$H sites,
we infer the evolution of the spectral functions $S_z(q,\omega_n)$
and $S_\perp(q,\omega_n)$. When the gap is significantly
reduced by the magnetic field, two different channels of nuclear
relaxation, specific to gapped antiferromagnets, are identified and
are in agreement with theoretical predictions.

\end{abstract}


]
\narrowtext

Several classes of one-dimensional Heisenberg antiferromagnets (HAF)
are known to exhibit a spin-gap at low temperature.
For example, integer-spin chains \cite{Haldane}
have a non-magnetic "spin-liquid" ground state (singlet) separated
from a branch of triplet excitations by an energy gap $\Delta$. A
spin-liquid ground state also exists in spin-ladders,
built by coupling an even number of $S=1/2$ HAF
chains with an antiferromagnetic transverse exchange $J_\perp$
\cite{Dagotto1,Dagotto2}. At low energies, many
physical properties are dominated by the singlet-triplet gap and do not
depend on the underlying dynamical quantum processes stabilizing the
ground state. For example, thermodynamic quantities
(susceptibility, specific heat) are very similar in a number of gapped
one-dimensional HAF.

In this letter, the low-energy dynamical processes dominating
the $^1$H spin-lattice relaxation rate ($1/T_1$) of an organic spin-ladder
(Cu$_2$(C$_5$H$_{12}$N$_2$)$_2$Cl$_4$)
are unambiguously identified by comparing the $T_1$ measurements
at two proton sites. They coincide precisely
with the processes proposed by Sagi and Affleck \cite{Sagi} for Haldane
systems. This experimental evidence supports the idea proposed by
Sachdev and coworkers \cite{Sachdev1} that spectral functions
$S_{z,\perp}(q, \omega)$ are, at low energies ($\omega \ll \Delta$),
common to all gapped one-dimensional HAF.

In Cu$_2$(C$_5$H$_{12}$N$_2$)$_2$Cl$_4$ \cite{Chiari},
the Cu$^{2+}$ (S=1/2) ions are coupled antiferromagnetically in
{\em well isolated} ladders \cite{LaCuO}
(see Fig. \ref{figstructure}).
The exchange parameters along the rungs ($J_\perp$) and the legs
($J_\|$) of the ladder are isotropic and have been accurately measured
to be $J_\perp=13.2$ K and $J_\|=2.4$ K \cite{Chabous,Hayward,Hammar1}.
In many respects, this material is a model system in which theoretical
predictions for Heisenberg ladders in the strong coupling limit
($J_\perp/J_\parallel \equiv 5.5 \gg 1$) can be tested.

The $^1$H NMR measurements were carried out by pulsed spin-echo
techniques on five single crystals (typically
$1\times 1\times 0.05$ mm$^3$), oriented with their $\hat b$ axis
(perpendicular to the chains axes) along the applied field
$H_0 = 5.6$ Tesla.
A typical spectrum for the proton resonance shows a
number of partially
resolved lines (Fig. 2a), indicating a variety of local fields
among the 24 inequivalent $^1$H sites. In the following, we focus
on the lines labelled (I) and (II), as their extreme position
in the spectrum allows their study on a wide temperature ($T$) range
\cite{commentlineI}.
With the field along $\hat b$, the magnetic hyperfine shift $K_{bb}$ of the
proton resonance is related to the uniform spin susceptibility
$\chi_0=\chi_{i}(q=0,\omega=0)$ at the nuclear site $i$, by
\begin{equation}
K_{bb}(T) = \frac{A_{bb}} {g_{bb}\mu_B} \chi_0(T) + \sigma,
\label{eq_shift}
\end{equation}
where $A_{bb}$ is the hyperfine coupling constant and $\sigma$
the chemical shift. The shift of the two lines, plotted in Fig. 2b,
are opposite in sign but follows
the same $T$-dependence as $\chi_0$, that is, a high temperature
Curie-Weiss behavior followed by an exponential drop below a rounded
maximum at $T_{\chi^{max}}\simeq$~8~K, in complete agreement with
previous susceptibility measurements\cite{Chabous,Hammar1}.
Since $K$ is proportional to the susceptibility $\chi_0$
measured at 5 T (Inset to Fig. 2b), the hyperfine
couplings on both sites can be estimated: $A_{bb}^{(I)}=+2.95\pm 0.40$ kOe and
$A_{bb}^{(II)}=-2.6\pm 0.50$ kOe \cite{commentsigma}. The largest
contribution to $A$ comes from the dipolar
field on the $j$-th nucleus created by the surrounding
electronic spins, i.e. $A_{j}\propto-|\gamma_e|\gamma_n\hbar^2\sum_{i}
{(1-3\cos^{2}\theta_{ij})/r^{3}_{ij}}$,
where $\theta_{ij}$ is the angle between $r_{ij}$ and $H_{0}$.
Given the atomic positions, it is straightforward to compute the dipolar
field at each $^1$H site (a reliable result is obtained by summing
over 5-6 neighboring Cu spins). The total NMR spectrum is
well-reproduced
in this way.  It is therefore possible to assign the NMR lines to
specific proton sites: line (II) is ascribed to protons
H2 involved in the superexchange $J_\parallel$
(See Fig. \ref{figstructure}).
The line (I) is attributed to protons H20 and H23 at the outer edges
of the ethyl groups. It must be stressed that the uncertainty
in the site labelling could only result in adding the protons H14
and H4 to the lines (I) and (II), respectively. This essentially
does not affect our analysis of the nuclear relaxation.

In a magnetic field, the triplet excitations split into three
branches. In the strong coupling limit ($J_\|/J_\perp \ll 1$),
the lowest branch is separated from the singlet ground state by
an effective gap $\Delta_h = \Delta - h$, where $h = g \mu_B H_0$
is the Zeeman energy.  When the temperature is small compared
to $\Delta_h$, interactions between excitations are negligible
and the lowest branch dominates the temperature dependence of the
susceptibility\cite{Troyer}
\begin{equation}
\chi_0 \propto \frac{1}{\sqrt{k_{\rm B}T J_\|}}
\exp \left( -\frac{\Delta_h}{k_{\rm B}T} \right).
\label{suscept}
\end{equation}
A low temperature fit of $K^{(I)}$ to Eq. (\ref{eq_shift}) and
(\ref{suscept}) gives an effective gap of $\Delta_h\simeq3$~K
in 5.6 Tesla.  This is very close to the value expected taking
the zero-field gap $\Delta=10.8$ K inferred from susceptibility
and high field magnetization measurement \cite{Chabous} reduced
by the Zeeman energy $h=7.6$ K.  Thus, the measurements of $K$
and $\chi_0$ are fully consistent with a spin gap
$\Delta_h=(J_\perp-J_\|) - h$ between the singlet and triplet states
of the Heisenberg ladder with strong rungs.

We now discuss the dynamical properties, as probed by the nuclear
spin-lattice relaxation rate $1/T_1$. The recovery of the nuclear
magnetization is always a single exponential at all temperatures.
As shown in Fig. 3a, both sites display qualitatively the same
$T$-dependence, that is, $1/T_1$ tends to be constant in the
paramagnetic region, and crosses over to an activated behavior
at low $T$. There are, however, striking differences between the two
lines:  (a) at high-$T$ the values of $T_1$ differ by one order of
magnitude, (b) at low-$T$ the gap values differ by a factor of 2.
Indeed, assuming an activated behavior
$1/T_1 \propto \exp(-\Delta_{eff}/k_{\rm B}T)$,
$\Delta_{eff}\simeq 3.4 \pm 0.2$ K for the line (I), close to the
value $\Delta_h=3.0$  K deduced from the shift, but
$\Delta_{eff}\simeq 6.8 \pm 0.2$ K for the line (II).

In spin systems, the temporal fluctuations of the electronic spins
make the nuclear polarization relax in a time $T_1$ related
to the spectral densities $S_{z,\perp}(q,\omega)$ of the two-spin
correlation functions, through \cite{Moriya}
\begin{equation}
\frac{1}{T_1}=\frac{(\gamma_n\gamma_e\hbar)^2}{2}
\sum_{q} [F_z S_z(q,\omega_n)+F_\perp S_\perp(q,\omega_n)],
\label{T1_def}
\end{equation}
with $\omega_n \sim 0$ the nuclear Larmor frequency, and
\begin{equation}
S_{z,\perp}(q,\omega_n)=
\int{dt e^{i\omega_n t}\langle \{S_{z,+}(q,t)S_{z,-}(-q,0)\} \rangle}.
\label{Sdeq}
\end{equation}
In general, any quantitative analysis of the relaxation requires
the knowledge of the hyperfine "form" factors $F(q)$ \cite{Henkens}
in addition to a model for $S_{z,\perp}(q,\omega)$. We first discuss the
structure factors $S_{z,\perp}(q,\omega)$.

Single magnon processes, which require an energy greater or equal to
the gap, cannot contribute to the nuclear relaxation which involves
negligible energy transfers $\hbar\omega_n\sim mK$.
Two- or three-magnon scattering processes are then required
\cite{Beeman}.  More specifically, Sagi and Affleck \cite{Sagi}
have recently analyzed the possible nuclear relaxation processes
for Haldane chains in magnetic fields.  Since the low energy
excitations of $S=1/2$ ladders and integer-spin chains are
qualitatively similar \cite{White}, it is natural to consider the same
processes here. Following their arguments, the nuclear spins can
exchange energy through three different channels \cite{Sagi}:\\
- $(i)$ {\it "intrabranch"} transitions involve two magnons
within the same branch (i.e. with the same $S_z$ eigenvalue).
At low $T$, these processes have a maximum probability near the minimum
at $k=\pi$ of the lowest branch of the triplet (Fig. \ref{figaffleck}),
implying a momentum transfer $\Delta k=q\sim 0$ (forward scattering).
For $T \ll h$, the $q$-integrated spectral density
is expected to follow the thermal occupation of the lowest energy
triplets $S^{intra}_z(\omega_n) \propto
\exp \left(-(\Delta-h)/k_{\rm B}T\right)$.\\
- $(ii)$ {\it "interbranch"} transitions (Fig. \ref{figaffleck}),
{\it i.e.} transitions from a state in a magnon branch $m$ to a state
with $m \pm 1$ ($S_\pm$ operators).  Since
the Zeeman splitting at 5.6 Tesla is larger than the magnon
bandwidth ($\sim5.5$ K \cite{Hammar2}), these processes can only occur
because of the finite damping of each level and are expected to be weak.
Furthermore, only large momentum transfers $q\sim\pi$
(backward scattering) remain at large Zeeman splitting. One infers from \cite{Sagi} that
$S^{inter}_\perp(\omega_n)\propto \exp(-\Delta/k_{\rm B}T)$.\\
- $(iii)$ {\it "staggered"} processes: when $H_0$ approaches
the critical field $h_{c1}\approx \Delta$, one-magnon excitations
($S_\pm$ operators, $q\sim\pi$) become increasingly relevant.
At finite $T$ in the gapped phase, interactions
between excitations, or equivalently  finite damping, generate
non-vanishing matrix elements for such transitions: this relaxation mechanism
involves three-magnon (or higher order) processes, and its
temperature dependence follows the square of the thermal population in
the lowest triplet state, $S^{stagg}_\perp(\omega_n)
\propto \exp \left(-2(\Delta-h)/(k_{\rm B}T) \right)$.

The above discussion shows that:
{\it (1)} the Boltzmann factor is more favorable to
intrabranch processes ($\Delta_h \simeq 3$ K); these will dominate the
staggered transitions ($2\Delta_h \simeq 6$ K), while interbranch ones,
if any, are essentially negligible ($\Delta \simeq 10$ K).
{\it (2)} The low-$T$ nuclear relaxation is only driven by two terms:
$S_z(q \sim 0,\omega_n)$ for intrabranch transitions and
$S_\perp(q\sim \pi,\omega_n)$ for staggered transitions. Accordingly,
we write Eq. (\ref{T1_def}) in a simplified form:
\begin{equation}
1/T_1 \propto F_z(0) S_z(q=0,\omega_n) +
F_\perp(\pi) S_\perp(q=\pi,\omega_n).
\label{T1simple}
\end{equation}

Hence, the two behaviors $1/T_1\propto
\exp (-\Delta_h/k_{\rm B} T)$ for the line (I) and $1/T_1\propto
\exp (-2\Delta_h/k_{\rm B} T)$ for (II) can only come from
the temperature dependence of $S_z(q=0, \omega_n) \propto \exp
(-\Delta_h/k_{\rm B}T)$ while
$S_\perp(q=\pi,\omega_n) \propto \exp (-2\Delta_h/k_{\rm B}T)$.
In other words, the ratio of $F_z(0)/F_\perp(\pi)$ for lines (I) and
(II) are such that, at low temperatures, only one of the exponential
terms dominates the relaxation: obviously, $S_z (q=0, \omega_{n})$
component is favored for the line (I),
and $S_{\perp} (q=\pi, \omega_{n})$ one for the line (II).

This result is, to our knowledge, the first experimental identification
of specific nuclear relaxation channels in a gapped antiferromagnet,
a result in remarkable agreement with the work of Sagi and Affleck.
Another support to this theory is that the lowest
gap $\Delta_h \simeq 3$ K is also the value seen in the
susceptibility.
Furthermore, the observation of the staggered contribution in the
gapped phase proves that interactions between fermionic-like
excitations are significant in this system.
This conclusion was already drawn from magnetization
measurements \cite{Chabous}.

A nice feature of this study is that the $T_1$ data for the two lines
provide a set of two independent equations ({\it i.e.} Eq.
\ref{T1simple} for each line).  Since $F_z(q)$ and $F_\perp(q)$ can be
computed for each
$^1$H site, the spectral functions $S_z$ and $S_\perp$ can,
in principle, be extracted separately.  Here, we found that
$F_\perp(q=\pi)$ is indeed five times
larger than $F_z(q=0)$ for the line (II),
while both terms are comparable for the line (I).
$S_\perp\propto \exp(-2\Delta_h/k_{\rm B}T)$ is thus {\it overweighted}
for the line (II) explaining why $T_1^{II}$ decays
with an activation energy $2 \Delta_{h}$.
On the other hand, $T_{1}^{I}$ is predominantly sensitive to the
smallest gap, $\Delta_{h}$, generated by $S_{z} (q=0, \omega_{n})$.
However, one must realize that the calculation of the form factors
is subject to several uncertainties:
any error in atomic positions is amplified
($F(r_{ij})\propto r^{-6}_{ij}$), the spatial extension of Cu$^{2+}$
orbitals may play an important role \cite{Renard} for the protons
H2 (line II) which are in the superexchange pathway
and closer to the Cu ion than those of line (I).
Indeed, the extracted values of $S_z$ are slightly negative suggesting
that the value of $F_\perp^{II}$ has been underestimated
in the calculation.  In fact, the pure Heisenberg paramagnetic
limit $S_z(q,\omega_n)=\frac{1}{2} S_\perp(q,\omega_n)$ should be
recovered when $T$ is large compared to $J_{\perp,\|}$ and $h$.
$F_\perp^{II}$ can be rescaled to a value satisfying
this limit at $T=30$ K, where the observed value of $T_1$
for the line (I) is within 10 \% of the paramagnetic limit calculated by
Moriya \cite{Moriya}.  In any event, this rescaling does not affect
the gap parameters extracted from the low-$T$ behaviour.
As shown in Fig. 3c, $S_\perp$ experiences a
gap $\Delta\simeq 6.8 \pm 0.2$~K {\em twice} as large as
in $S_z$ ($3.4 \pm 0.2$~K).

This analysis of the nuclear relaxation in
Cu$_2$(C$_5$H$_{12}$N$_2$)$_2$Cl$_4$, a strongly coupled ladder,
settles an ongoing controversy:
the gap values derived from static
($\chi_0$) and dynamic $S_z(q\sim 0,\omega_n)$ measurements are
here {\em identical}, in complete agreement with the predictions
of Troyer {\it et al.} \cite{Troyer} and Sagi and Affleck \cite{Sagi}.
This strongly contrasts with the experimental observations in inorganic
ladders \cite{NMRSrCuO,VO2P2O7}, and in some Haldane
chains \cite{Shimizu,Takigawa}.
In these materials, the different temperature dependence
observed in the dynamics may be due to a second minimum
in the dispersion relation. For example, it should be the case in
$\rm SrCu_{2}O_{3}$ if $J_\perp < J_\|$ \cite{Oitmaa96,Johnston}.
Low-lying excitations near $k=0$ such that
$\Delta_{k=0} \sim \Delta_{k=\pi}$ may open up relaxation
channels involving large-$q$ interbranch transitions
and would lead to a higher effective gap in $T_1$ measurements.
Other explanations have been proposed in the
limits $T \ll \Delta_h$ \cite{Sachdev2} and
$J_\perp \ll J_\|$ \cite{Fukuyama}.

In conclusion, $^1$H NMR experiments demonstrate that the effective
gap of the $S=1/2$ HAF ladder Cu$_2$(C$_5$H$_{12}$N$_2$)$_2$Cl$_4$ in
a magnetic field $h$ is $\Delta_h \simeq J_\perp-J_\|-h$.
The nuclear relaxation can be quantitatively understood
in the framework of the theory of Sagi and Affleck, retaining only
"intrabranch" and "staggered" processes. More generally, the processes
identified in this work should be generic to many
gapped HAF chains in a magnetic field.

We thank J.-P. Boucher, K. Damle and S. Sachdev for useful discussions. The GHMFL is
Laboratoire Conventionn\'e aux Universit\'es J. Fourier et INPG
Grenoble I.

\begin{figure}
\caption{Schematic structure of Cu$_2$(C$_5$H$_{12}$N$_2$)$_2$Cl$_4$,
with the exchange parameters determined in Ref. \protect\cite{Chabous}.
The labelled protons contribute to the two NMR lines used
in this work to probe different dynamical functions
($S_z$ and $S_\perp$, see text).}
\label{figstructure}
\end{figure}
\begin{figure}
\caption{(a): $^1$H NMR spectra at fixed frequency $f_0=239.112$ MHz.
The arrows indicate the two lines studied (their different amplitudes
are due to different excitation conditions).  The large central peak
comes from protons in the NMR probe.
(b): magnetic hyperfine shift for the lines ($I$) and ($II$)
and susceptibility at 5 Tesla; the dashed line is a fit where the
hyperfine coupling is the only adjustable parameter and
temperature dependence of the susceptibility is given by Eq. 5 of Ref.
\protect\cite{Chabous}, with $J_\perp=13.2$ K and $J_\|=2.5$ K.
Inset: shift data vs. susceptibility, with $T$ as an implicit
parameter.}
\end{figure}
\begin{figure}
\caption{$^1$H spin-lattice relaxation rate $1/T_1$ for lines (I)
and (II) as a function of $T$ (a) and $1/T$ (b).
(c): $S_{z} (q=0, \omega_{n})$ and $S_{\perp} (q = \pi, \omega_{n})$
correlation functions derived from $T_1$ results
(see text for details).}
\end{figure}
\begin{figure}
\caption{Schematic picture of the two-magnon scattering processes
relevant to the nuclear relaxation in a system with singlet to triplet
gap, in a magnetic field $H_0$ \protect\cite{Sagi}.
In this experiment, the Zeeman splitting $g\mu_BH
\sim7.6$ K is larger than the magnon bandwidth
($\sim$5.5 K \protect\cite{Hammar2}).}
\label{figaffleck}
\end{figure}

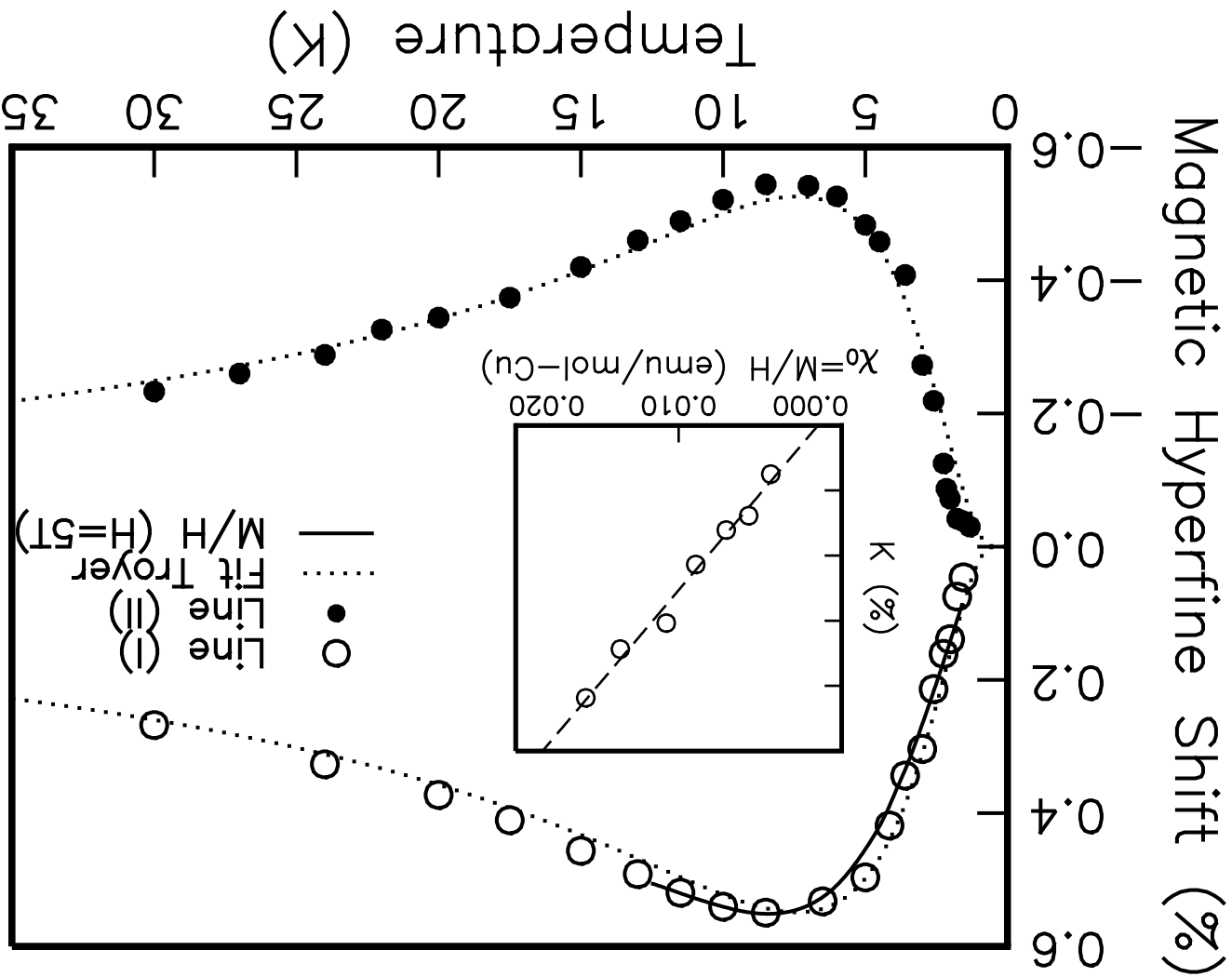
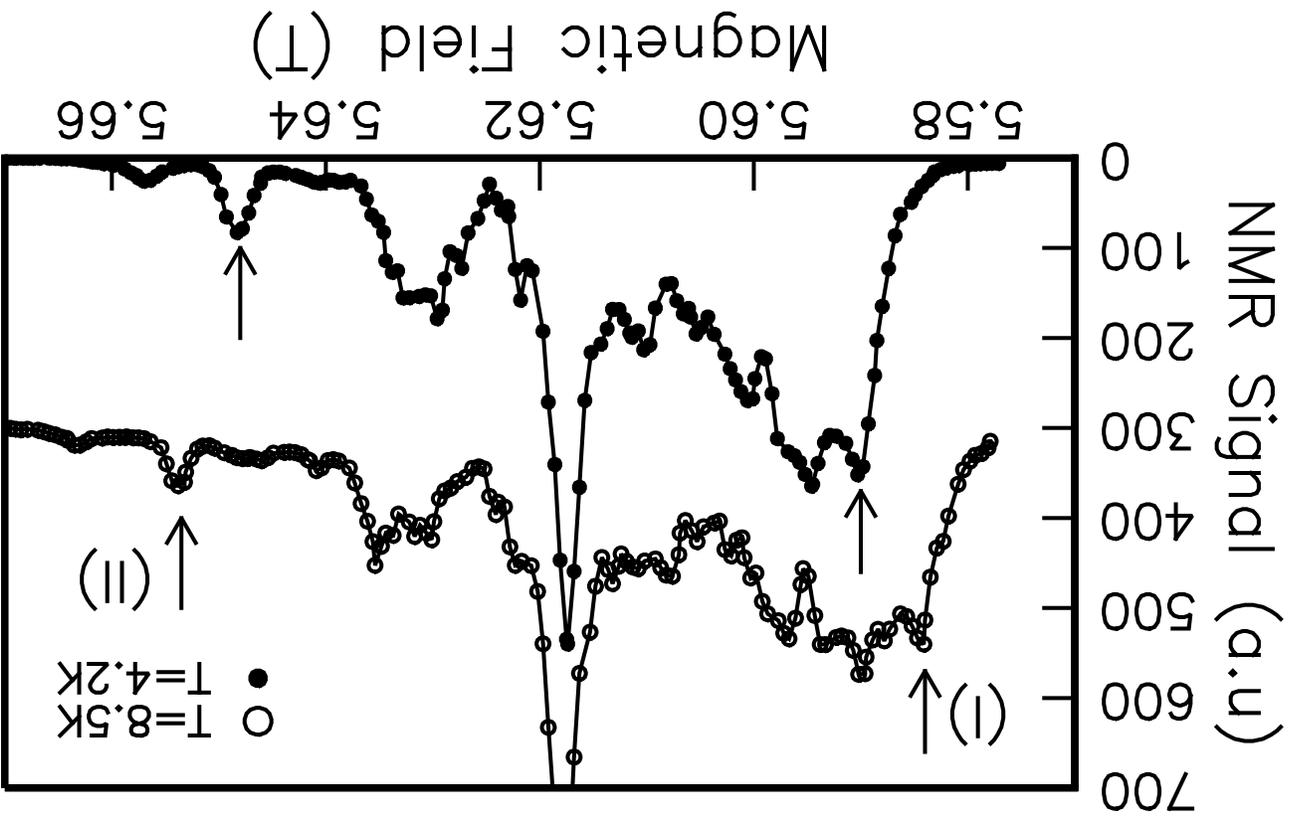
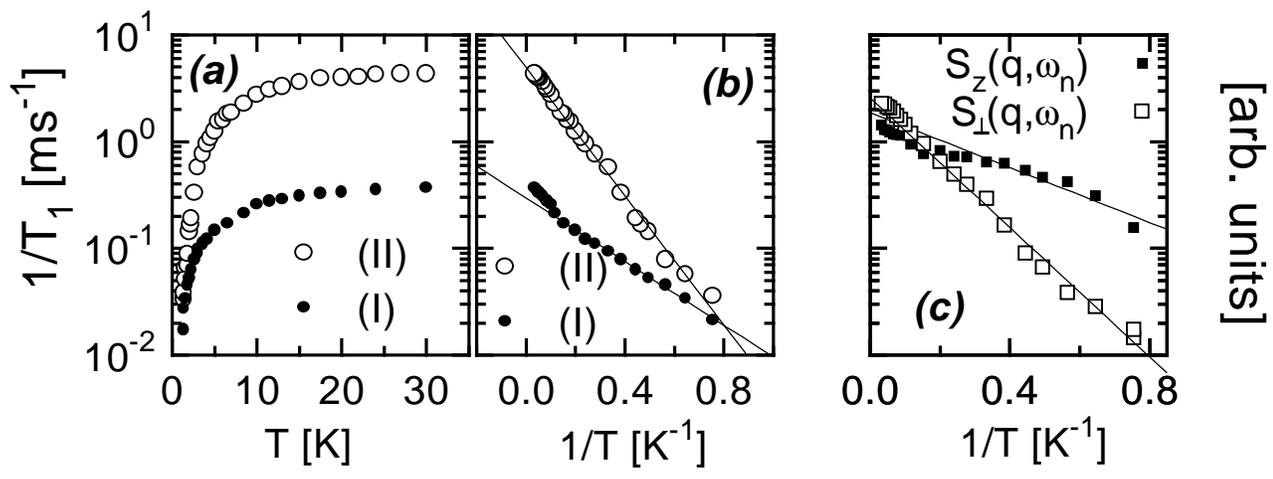
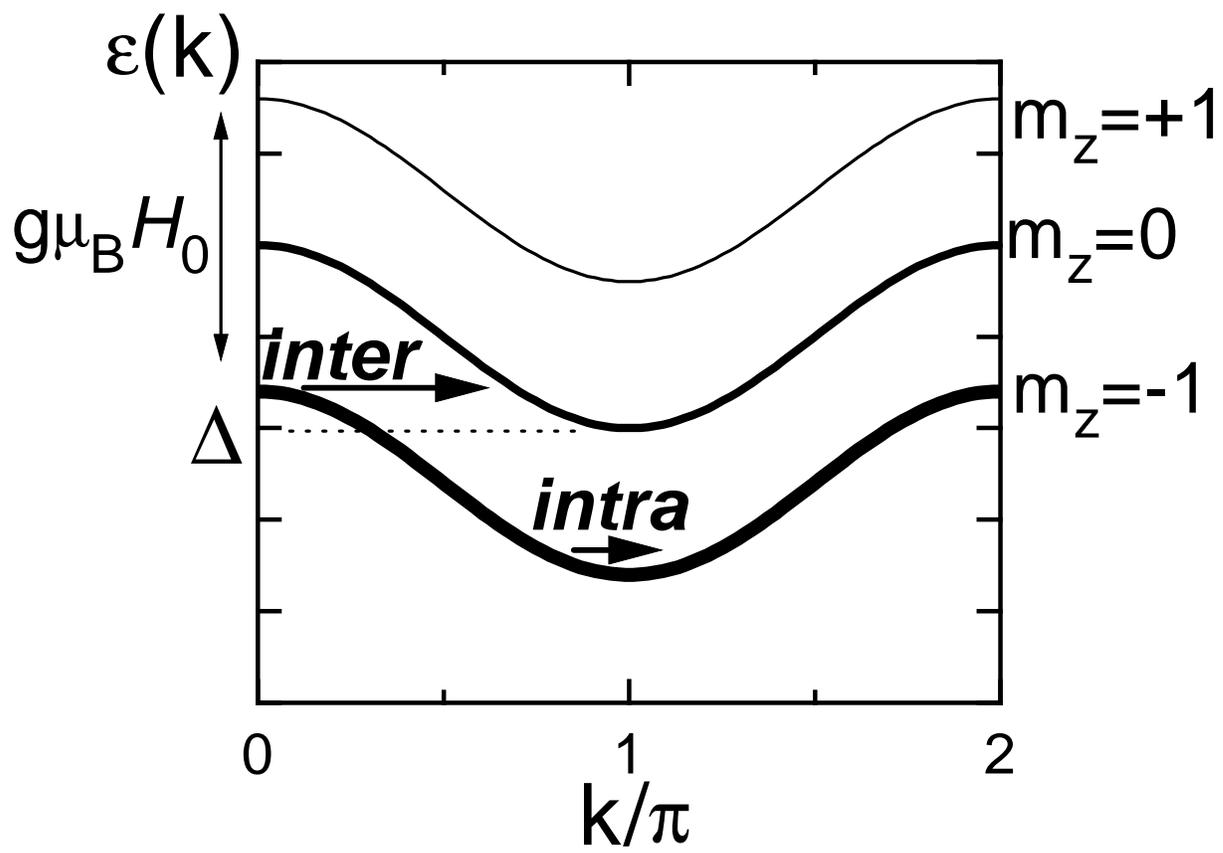
\end{document}